\documentclass[twocolumn,showpacs,preprintnumbers,amsmath,amssymb,epsfig,widetext]{revtex4}
\usepackage{graphicx}
\usepackage{dcolumn}
\usepackage{bm}
\usepackage{epsfig}
\usepackage{color}
\usepackage{ulem}
\usepackage{hyperref}

\hypersetup{
    colorlinks=true,
    linkcolor=red,
    citecolor=blue,
} 

\input{epsf}
\newcommand{\beq}{\begin{equation}}
\newcommand{\eeq}{\end{equation}}
\newcommand{\bea}{\begin{eqnarray}}
\newcommand{\eea}{\end{eqnarray}}
\newcommand{\bi}{\begin{itemize}}
\newcommand{\ei}{\end{itemize}}
\newcommand{\bfi}{\begin{figure}[!t]
\epsfxsize=7cm
\epsffile}
\newcommand{\bfib}{\begin{figure}[htb]
\epsfxsize=9cm
\epsffile}
\newcommand{\bfig}{\begin{figure*}[htb]
\epsfxsize=12cm
\epsffile}
\newcommand{\efi}{\end{figure}}
\newcommand{\efib}{\end{figure}}
\newcommand{\efig}{\end{figure*}}
\newcommand{\no}{\nonumber}

\newcommand{\aj}{AJ}

\newcommand{\mnras}{MNRAS}
\newcommand{\jcap}{JCAP}
\newcommand{\physrep}{Physics reports}
\newcommand{\na}{New Astronomy}

\newcommand{\araa}{Ann. Rev. Astron. Astrophys.}

\newcommand{\bfs}{\mbox{\boldmath$s$}}
\newcommand{\bfr}{\mbox{\boldmath$r$}}
\newcommand{\bfx}{\mbox{\boldmath$x$}}
\newcommand{\bfk}{\mbox{\boldmath$k$}}
\newcommand{\bfv}{\mbox{\boldmath$v$}}
\newcommand{\bfu}{\mbox{\boldmath$u$}}

\newcommand{\hompc}{\,h\,{\rm Mpc}^{-1}}

\newcommand{\mpcoh}{\,h^{-1}\,{\rm Mpc}}

\def\be{\begin{equation}}
\def\ee{\end{equation}}
\def\ba{\begin{eqnarray}}
\def\ea{\end{eqnarray}}

\def\nn{\nonumber}

\begin{document}

\preprint{YITP-21-09}

\title{Toward a more stringent test of gravity with redshift space power spectrum: simultaneous probe of growth and amplitude of large-scale structure}
 
\author{Yong-Seon Song$^{1}$, Yi Zheng$^{2}$, Atsushi Taruya$^{3,4}$}
\email{ysong@kasi.re.kr}
\affiliation{$^{1}$ Korea Astronomy and Space Science Institute, Daejeon 34055, Republic of Korea \\
$^2$ School of Physics and Astronomy, Sun Yat-sen University, 2 Daxue Road, Tangjia, Zhuhai, 519082, China\\
$^3$ Center for Gravitational Physics, Yukawa Institute for Theoretical Physics, Kyoto University, Kyoto 606-8502, Japan,\\
$^4$ Kavli Institute for the Physics and Mathematics of the Universe, Todai Institutes for Advanced Study, the University of Tokyo, Kashiwa, Chiba 277-8583, Japan (Kavli IPMU, WPI) 
}
\date{\today}

\begin{abstract}
Redshift-space distortions (RSD) offers an exciting opportunity to test the gravity on cosmological scales. In the presence of galaxy bias, however, the RSD measurement at large scales, where the linear theory prediction is safely applied, is known to exhibit a degeneracy between the parameters of structure growth $f$ and fluctuation amplitude $\sigma_8$, and one can only constrain the parameters in the form of $f\sigma_8$. In order to disentangle this degeneracy, in this paper, we go beyond the linear theory, and consider the model of RSD applicable to a weakly nonlinear regime. Based on the Fisher matrix analysis, we show explicitly that the degeneracy of the parameter $f\sigma_8$ can be broken, and $\sigma_8$ is separately estimated in the presence of galaxy bias. Performing further the Markov chain Monte Carlo analysis, we verify that our model correctly reproduces the fiducial values of $f\sigma_8$ and $\sigma_8$, with the statistical errors consistent with those estimated from the Fisher matrix analysis. We show that upcoming galaxy survey of the stage-IV class can unambiguously determine $\sigma_8$ at the precision down to $\lesssim10$\% at higher redshifts even if we restrict the accessible scales to $k\lesssim0.16\,h$\,Mpc$^{-1}$. 
\end{abstract}


 
\maketitle

\section{Introduction} 
\label{sec:intro}

Since its discovery two decades ago~\cite{Acceleration1,Acceleration2}, the origin of late-time cosmic acceleration has remained puzzled. While the flat Lambda cold dark matter ($\Lambda$CDM) model is currently the best as concordant cosmological model to describe both the cosmic expansion and structure formation, consistent with observations of cosmic microwave background and large-scale structure, the tension with the cosmological parameters determined at the local universe has been recently highlighted, suggesting a need of new physics beyond $\Lambda$CDM model~\cite{Riess2016,H0LiCOW2020,Jedamzik2020}.

Theoretically, the origin of cosmic acceleration can be explained by either the presence of a mysterious energy component called dark energy or a long-distance modification of gravity, referred to as modified gravity ~\cite{Lue2006,Frieman2008,Li2011,Clifton2012,Joyce2016,Koyama2016,Nojiri2017,Arun2017,Wang2017,Brax2018,Mustapha18}.  In order to realize the accelerated cosmic expansion, the former introduces a negative pressure support, and the latter changes a law of gravitational physics on large scales. Observationally discriminating between two scenarios therefore requires a simultaneous measurements of the cosmic expansion and growth of structure.

Among various cosmological probes, the redshift-space galaxy clustering offers a sensible probe of both the cosmic expansion and growth of structure. At large scales, the baryon acoustic oscillations (BAO) imprinted on the clustering pattern of galaxies appears statistically manifest, and it can be used for a standard ruler to determine the angular diameter distance ($D_A$) and Hubble parameters $H$ at high redshifts through the Alcock-Paczynski test \cite{APtest}. Further, the observed galaxy distribution via spectroscopic surveys is statistically anisotropic due to the peculiar velocity of galaxies by Doppler effect, referred to as the redshift-space distortions (RSD). In linear theory, the strength of anisotropies is solely characterized by the growth rate $f$, defined by $f=d\ln D_+/d\ln a$, with $D_+$ and $a$ being linear growth factor and scale factor of the Universe, respectively. Thus, through a precision measurement of redshift-space galaxy power spectrum or correlation function, one can in principle obtain simultaneously the information on the three parameters (i.e., $D_A$, $H$, and $f$). 

However, in the presence of the galaxy bias, the situation becomes bit complicated. To be precise, consider the large scales where the linear theory is safely applied. Then, the galaxy power spectrum is generally modeled as
\beq
P^{(S)}_g(k,\mu) = \left[b_1\sigma_8(z)+f\sigma_8(z)\mu^2\right]^2 P_{\rm m}(k)\,,
\label{eq:linear_pk}
\eeq
where the variable $\mu$ is the directional cosine between wavevector and line-of-sight direction. The function $P_{\rm m}$ is the matter power spectrum normalized at the present day. The quantity $\sigma_8(z)$ is the root-mean-square mass fluctuation in spheres with radius $8\mpcoh$ at redshift $z$, and is recast in linear theory as $\sigma_8(z)=\sigma_8(0)D_+(z)$ with $D_+$ being normalized to unity at $z=0$. The $b_1$ is the linear bias parameter. Note that taking further the Alcock-Paczynski effect into account, the projected wavenumbers perpendicular and parallel to the line-of-sight direction, $k_\perp$ and $k_\parallel=\mu\,k$, are respectively replaced with $D_A/D_{A,{\rm fid}}k_\perp$ and $(H/H_{\rm fid})^{-1}k_\parallel$, and the power spectrum given above is further multiplied by $(H/H_{\rm fid})(D_A/D_{A,{\rm fid}})^{-2}$, where the quantities with subscript indicate those estimated in a fiducial cosmological model.

The structure of Eq.~(\ref{eq:linear_pk}) indicates that the parameter $\sigma_8$ is degenerated with growth rate $f$ and bias $b_1$, and one can only determine the combinations of the parameters, i.e., $b_1\sigma_8$ and $f\sigma_8$, through the observed power spectrum. In other words, unless the bias parameter $b_1$ is known a priori, one cannot break the degeneracy between growth rate $f$ and $\sigma_8$. Note that the Alcock-Paczynski effect induces distinctive anisotropies in the measured power spectrum, and making use of BAO features, one can separately determine $D_A/D_{A,{\rm fid}}$ and $H/H_{\rm fid}$ without any degeneracy with $\sigma_8$.

Toward a solid test of gravity and cosmic acceleration, the degeneracy between $f$ and $\sigma_8$ has to be broken. To do this, one simple approach is to combine the power spectrum with other cosmological probe. In this respect, the use of galaxy bispectrum would be obviously important, and this can also provide an additional cosmological information, further tightening the cosmological constraints~\cite{Song15c,Hector_bias}. Another approach is to stick to the power spectrum, and to use the small-scale information beyond the linear regime. Recalling that Eq.~(\ref{eq:linear_pk}) is valid only at large scales, if we go to nonlinear regime, there appear corrections involving the parameters $\sigma_8$ and $f$ but with a different combination. In fact, using the perturbation theory calculation, we can identify the directional-dependent terms proportional to $f^2\sigma_8^4$ or $f^3\sigma_8^4$ in the matter power spectrum \cite{Taruya10,Taruya13}. Thus, provided an accurate theoretical model, accessing (weakly) nonlinear scales would give a way to break the degeneracy between $f$ and $\sigma_8$. A price to pay is, however, the new degrees of freedom to characterize the galaxy bias at nonlinear scales. That is, beyond linear scales, we need to introduce several bias parameters describing the nonlinear modification to power spectrum, and one has to marginalize them to determine parameters sensitive to the cosmology. It is thus not trivial at all that the growth rate $f$ is uniquely determined without any degeneracy.

In this paper, based on a nonlinear theoretical model of the redshift-space galaxy power spectrum, we explicitly demonstrate that the degeneracy between $f$ and $\sigma_8$ is broken at weakly nonlinear scales. The model specifically considered here is called the hybrid RSD model that has been developed by Ref.~\cite{Zheng16a,Song2018,Zheng19}. This is a perturbation theory based model \cite{Taruya10,Taruya13}, but partly including the correction terms calibrated by $N$-body simulations, the accuracy is improved and the applicable range is extended \cite{Song2018}. In this paper, incorporating further the nonlinear bias prescription into the hybrid RSD model, we will investigate how the degeneracy between $f$ and $\sigma_8$ is broken, and quantify the expected constraints on parameters $\sigma_8$, $f\sigma_8$ as well as $D_A$ and $H$ for a representative galaxy survey (Dark Energy Spectroscopy Instrument, DESI).

Related to the present study, one may comment on the full-shape analysis that recently performed using SDSS BOSS galaxies~\cite{Ivanov:2019pdj,DAmico:2019fhj}. In this study, based on the effective field theory of large-scale structure, the input linear power spectrum is allowed to vary, enabling us to directly constrain each cosmological parameters. Going beyond linear regime, not only tightening constraints but also breaking parameter degeneracy is shown to be manifest~\cite{Nishimichi:2020tvu},  assuming the underlying theory of gravity. The spirit of this approach is close to the present paper, but we here stick to the consistent test of gravity, and take $f$ or $f\sigma_8$ to be a free parameter,  independent of other cosmological parameters.

This paper is organized as follows. Sec.~\ref{sec:hybrid_RSD_model} presents a model of the redshift-space galaxy power spectrum applicable to the weakly nonlinear regime. Based on this model as a theoretical template, in Sec.~\ref{sec:Fisher_forecast}, we use the Fisher matrix formalism to quantitatively investigate how well one can break the degeneracy of the parameters $f\sigma_8$. Sec.~\ref{sec:test_MCMC} examines the Markov chain Monte Carlo analysis to estimate the cosmological parameters in the hal catalog. We verify that our model of RSD faithfully reproduces the fiducial parameters in the $N$-body simulations, consistent with the Fisher forecast results. Finally, Sec.~\ref{sec:conclusion} is devoted to conclusion and discussion.

\section{Hybrid RSD model for galaxy clustering} 
\label{sec:hybrid_RSD_model} 


In this section, we present the model of redshift-space galaxy power spectrum beyond the linear regime. After briefly reviewing the hybrid RSD model for matter fluctuations in Sec.~\ref{subsec:hybrid_DM_RSD}, we extend it to incorporate the nonlinear galaxy bias in Sec.~\ref{subsec:modeling_galaxy_bias}.

\subsection{Hybrid RSD model for matter power spectrum}
\label{subsec:hybrid_DM_RSD}

In order to model the observed galaxy power spectrum, one critical ingredient is the redshift-space distortions (RSD). In principle, all the effects of RSD is accounted for by the simple relation between real-space ($\bfr$) and redshift-space ($\bfs$) positions: 
\beq
\label{eq:mapping}
\bfs=\bfr+\frac{\bfv \cdot \hat{z}}{aH}\hat{z},
\eeq
where the quantities $\bfv$, $a$ and $H$ respectively denote the physical peculiar velocity, the scale factor of the Universe, and the Hubble parameter. In this paper, we will take the distant-observer limit, and choose the $z$-direction as the line-of-sight direction. With the mapping relation given above, the power spectrum in redshift space is generally expressed in terms of the real-space quantities (e.g., Ref.~~\cite{Taruya10}):   
\begin{equation}
P^{\rm(S)}(k,\mu)=\int d^3\bfx\,e^{i\,\bfk\cdot\bfx}
\bigl\langle e^{j_1A_1}A_2A_3\bigr\rangle\,, 
\label{eq:Pkred_exact}
\end{equation}
where the variable $j_1$ and functions $A_i$ are defined as follows:
\begin{eqnarray}
&j_1= -i\,k\mu ,\nonumber\\
&A_1=u_z(\bfr)-u_z(\bfr'),\nonumber\\
&A_2=\delta(\bfr)+\,\nabla_zu_z(\bfr),\nonumber\\
&A_3=\delta(\bfr')+\,\nabla_zu_z(\bfr').\nonumber
\end{eqnarray}
The quantities $\bfx$ and  $\bfu$ are given respectively by $\bfx=\bfr-\bfr'$ and $\bfu\equiv-\bfv/(aH)$. The function $u_z$ is the line-of-sight  component of $\bfu$. 

Based on Eq.~(\ref{eq:Pkred_exact}), a rigorous calculation of the redshift-space power spectrum generally requires a non-perturbative treatment. This is true even if the density and velocity follow the Gaussian statistics. We thus employ the perturbative treatment and derive the expression relevant to the weakly nonlinear regime. To do this, one proposition made in Refs.~\cite{Taruya10,Zheng14a} is that a part of the zero-lag correlation in the exponent is kept as a non-perturbative contribution, while rest of the terms is Taylor-expanded. The resultant expression for the power spectrum, relevant at the next-to-next-to-leading order becomes \cite{Taruya10, Taruya13,Zheng14a}
\begin{align}
\label{eq:Pkred_final}
P^{\rm (S)}(k,\mu)
&=D^{\rm FoG}(k\mu\sigma_z)\,\Bigl[P_{\delta\delta}(k)+2\mu^2P_{\delta\Theta}(k)+\mu^4P_{\Theta\Theta}(k)  
\nn
\\
&\qquad
+A(k,\mu)+B(k,\mu)+T(k,\mu)+F(k,\mu)\Bigr] \nonumber,
\end{align}
where the quantity $\Theta$ is the velocity-divergence field, $\Theta\equiv -\nabla\cdot \bfv/(aH)=\nabla \cdot \bfu$. The spectra, $P_{\delta\delta}$, $P_{\delta\Theta}$, and $P_{\Theta\Theta}$ are respectively the auto-power spectrum of density, velocity-divergence fields, and their cross-power spectrum. The first line in the bracket is originated from the term $\langle A_2A_3\rangle_c$, and is obtained assuming the irrotational flow. At the linear order, it is reduced to the squashing Kaiser term, i.e., Eq.~(\ref{eq:linear_pk}). The rest of the terms in the bracket are the higher-order corrections characterizing the nonlinear correlations between the density and velocity fields, defined by
\begin{eqnarray}
  A(k,\mu)&=& j_1\,\int d^3\bfx \,\,e^{i\bfk\cdot\bfx}\,\,\langle A_1A_2A_3\rangle_c,\nonumber\\
  B(k,\mu)&=& j_1^2\,\int d^3\bfx \,\,e^{i\bfk\cdot\bfx}\,\,\langle A_1A_2\rangle_c\,\langle A_1A_3\rangle_c,\nonumber\\
  T(k,\mu)&=& \frac{1}{2} j_1^2\,\int d^3\bfx \,\,e^{i\bfk\cdot\bfx}\,\,\langle A_1^2A_2A_3\rangle_c,\nonumber \\
  F(k,\mu)&=& -j_1^2\,\int d^3\bfx \,\,e^{i\bfk\cdot\bfx}\,\,\langle u_z u_z'\rangle_c\langle A_2A_3\rangle_c. \nonumber
\end{eqnarray}
Note that the factorized term $D^{\rm FoG}$ in Eq.~(\ref{eq:Pkred_final}) represents a non-perturbative contribution coming from the zero-lag correlation of velocity fields, and it plays a role to suppress the overall amplitude at small scales. The explicit functional form will be later specified (see Sec.~\ref{sec:Fisher_forecast}).

To explicitly compute Eq.~(\ref{eq:Pkred_final}) in the case of the matter fluctuations, in Ref.~\cite{Song2018}, we have used the re-summed perturbation theory treatment by the multi-point propagator expansion \cite{RegPT,Taruya13,Bernardeau2008}. To account further for a small but non-negligible flaw in the perturbative calculations, we have added the corrections calibrated by the $N$-body simulations, which enables us to predict the redshift-space matter power spectrum at the precision of $\lesssim 1$\% down to $k\simeq0.18\hompc$. In Appendix \ref{appendix:hybrid}, the description of our hybrid model is presented for the matter power spectrum, and the explicit dependence on the quantities $\sigma_8(z)$ and $f(z)$ are shown.

Although the expressions of the higher-order contributions are rather intricate, these terms involve the contributions expressed in the polynomial form of $f^m\sigma^n$, with the power-law indices $m$ and $n$ running respectively from $1$ to $4$ and $4$ to $6$ for the perturbative calculations at next-to-next-to-leading order. Besides, beyond the linear order, even the auto-power spectra $P_{\delta\delta}$ and $P_{\Theta\Theta}$ do not simply scale as $\sigma_8^2$ and $(f\,\sigma_8)^2$, respectively. Thus, if one gets access to the weakly nonlinear regime where the higher-order terms play a role, we expect that the degeneracy of the parameter $f\sigma_8$ at Eq.~(\ref{eq:linear_pk}) is broken, and $f$ and $\sigma_8$ can be separately determined.


\begin{figure*}
\begin{center}
\resizebox{3.in}{!}{\includegraphics{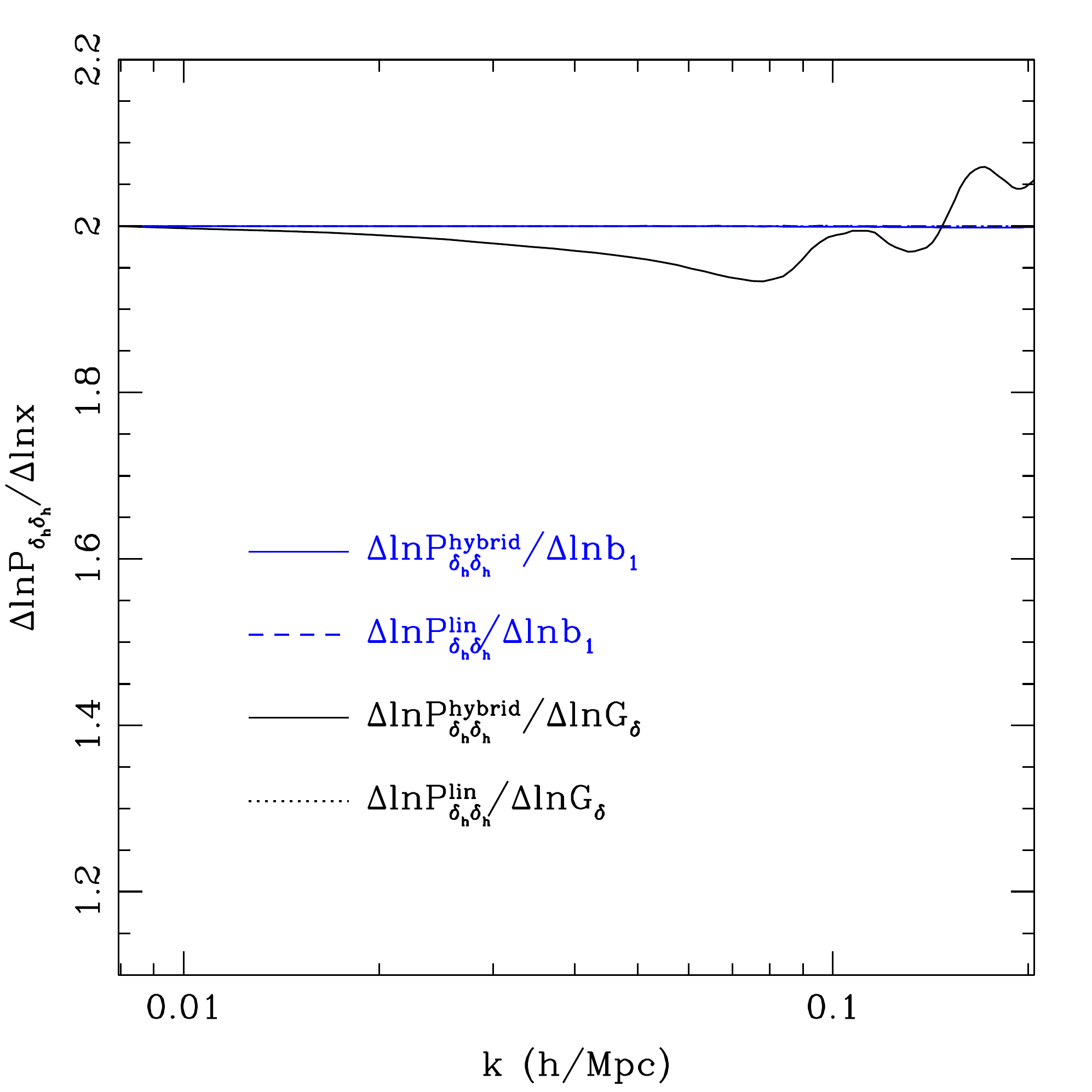}}\hfill
\resizebox{3.in}{!}{\includegraphics{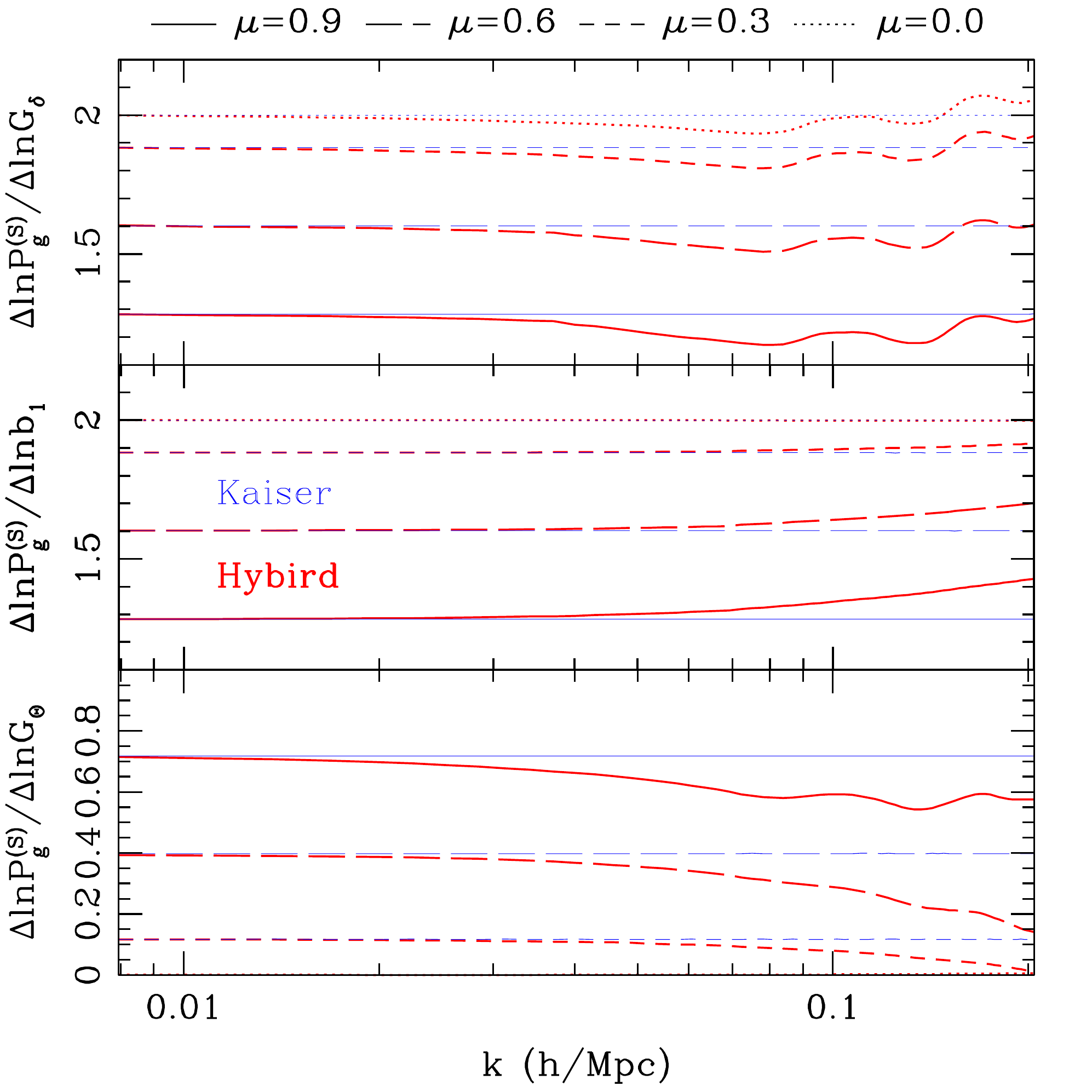}} 
\vspace*{0.5cm}
\end{center}
\caption{{\it Left}: Logarithmic derivative of the real-space halo power spectrum with respect to the linear growth factor $G_\delta$ (black) and linear bias parameter $b_1$ (blue), plotted as function of the wavenumber $k$. In each panel, the solid lines represent the results for hybrid model, $P_{\delta\delta}^{\rm hybrid}$, while dashed and dotted lines are for the linear theory prediction,  $P_{\delta\delta}^{\rm lin}$. {\it Right}: logarithmic derivative of the redshift-space halo power spectra with respect to $G_\delta$, $b_1$ and $G_\Theta$, respectively shown from top to bottom panels. Both of the Kaiser (linear theory) and hybrid models are shown as blue and red curves, specifically fixing the directional cosine to $\mu=0.0$ (dotted), $0.3$ (short-dashed), $0.6$ (long-dashed) and $0.9$ (solid). }
\label{fig:dPdx}
\end{figure*}

\subsection{Modeling galaxy bias}
\label{subsec:modeling_galaxy_bias}

On top of the hybrid RSD model in previous subsection, there is one more step toward a practical application to the observed galaxy power spectrum. Since the galaxy distribution is a biased tracer of matter distribution, accounting for the galaxy bias is another crucial task. In order to do this, one may recall that the expression given at Eq.~(\ref{eq:Pkred_final}) is fairly general. Replacing the matter density field $\delta$ with the galaxy/halo density field $\delta_h$, Eq.~(\ref{eq:Pkred_final}) can be applied to the observed power spectrum, and hence our hybrid model of matter power spectrum is used as a building block to compute accurately the redshift-space galaxy power spectrum.

For our interest at the weakly nonlinear regime, a perturbative description of the tracer field is valid, and $\delta_h$ is expanded in powers of the matter density field $\delta$, including the non-local contributions. Here, we adopt the prescription proposed by Ref.~\cite{McDonald_bias,Saito2014}, which has been applied to the SDSS BOSS (e.g., \cite{Hector_bias}) and eBOSS galaxies \cite{Arnaud2020}. Apart from the stochastic terms, this is a general perturbative expansion valid at the next-to-leading order. While the hybrid model for the matter power spectrum in Sec.~\ref{subsec:hybrid_DM_RSD} and Appendix \ref{appendix:hybrid} includes the corrections valid at next-to-next-to-leading order, we shall below consider the scales where the next-to-leading order is important, but the next-to-next-to-leading order is still subdominant. A fully consistent treatment including the bias at next-to-next-to-leading order will be discussed in our future work.

Based on Refs.~\cite{McDonald_bias,Saito2014}, the auto-power spectrum of the tracer density field, $P_{\tilde{\delta}_h\tilde{\delta}_h}$, is expressed as follows:
\bea
P_{\tilde{\delta}_h\tilde{\delta}_h}(k)&=&P_{\delta_h\delta_h}(k)+P_{\epsilon\epsilon}\,,
\label{eq:pdd_bias}
\eea
where the term $P_{\epsilon\epsilon}$ is the stochastic contribution mainly characterizing the shot noise, which is usually constant, and is estimated, assuming the Poisson noise, from the number density of tracer field. The first term, $P_{\delta_h\delta_h}$, is the deterministic part, and is given in the following form: 
\begin{align}
P_{\delta_h\delta_h}(k)&=b_1^2P_{\delta\delta}(k)+2b_1b_2P_{b2,\delta}(k)+2b_2b_{s2}P_{bs2,\delta}(k) 
\no
 \\
&
+2b_1b_{3\rm{nl}}\sigma_3^2(k)P^{\rm{L}}_{\rm m}(k)+b_2^2P_{b22}(k)
\no
 \\
&
+2b_2b_{s2}P_{b2s2}(k)
+b_{s2}^2P_{b22}(k). 
\end{align}
On the other hand, the cross-power spectrum, $P_{\delta_h\theta_h}$, contains only the deterministic contribution, whose expression is given by
\begin{align}
P_{\delta_h\Theta_h}(k)&=b_1P_{\delta\Theta}(k)+b_2P_{b2,\Theta}(k)+b_{s2}P_{bs2,\Theta}(k)\no
\\
&+b_{3\rm{nl}}\sigma_3^2(k)P^{\rm{L}}_{\rm m}(k)\,.
\label{eq:pdt_bias}
\end{align}
Note that in the absence of velocity bias, the auto-power spectrum of galaxy/halo velocity field, $P_{\Theta_h\Theta_h}$, is identical to $P_{\Theta\Theta}$. In Eqs.~(\ref{eq:pdd_bias}) and (\ref{eq:pdt_bias}), while the first terms, $b_1^2P_{\delta\delta}$ and $b_1P_{\delta\Theta}$, as well as $P_{\Theta_h\Theta_h}$, are the leading-order bias contribution, and can be computed with the hybrid RSD model for the matter power spectrum, the terms involving parameters $b_2$, $b_{s2}$, and $b_{\rm 3nl}$ are the higher-order bias terms. The explicit expressions for their scale-dependent functions, $P_{b2,\delta}$, $P_{bs2,\delta}$, $\sigma_3^2$, $P_{b22}$, ..., can be found in Ref.~\cite{McDonald_bias,Saito2014}.

Regarding the $A$, $B$, $T$ and $F$ terms in Eq.~(\ref{eq:Pkred_final}), they are all regarded as the higher-order corrections. We only apply the linear bias $b_1$ to these terms \cite{Zheng19},
\bea
\label{eq:higherorder_b1_A}
A_h(k,\mu)&=&b_1^3A(k,\mu, f/b_1)\,,\\
\label{eq:higherorder_b1_B}
B_h(k,\mu)&=&b_1^4B(k,\mu,f/b_1)\,, \\
\label{eq:higherorder_b1_F}
F_h(k,\mu)&=&b_1^4F(k,\mu,f/b_1)\,, \\
T_h(k,\mu)&=&b_1^4T(k,\mu,f/b_1)\,,
\label{eq:higherorder_b1_T}
\eea
where all the subscripts $h$ at left-hand-side represent the higher-order terms of the tracer fields. The accuracy of this treatment and its impacts on the RSD model was discussed in Ref.~\cite{Zheng19}.

\begin{figure*}
\begin{center}
\resizebox{3.in}{!}{\includegraphics{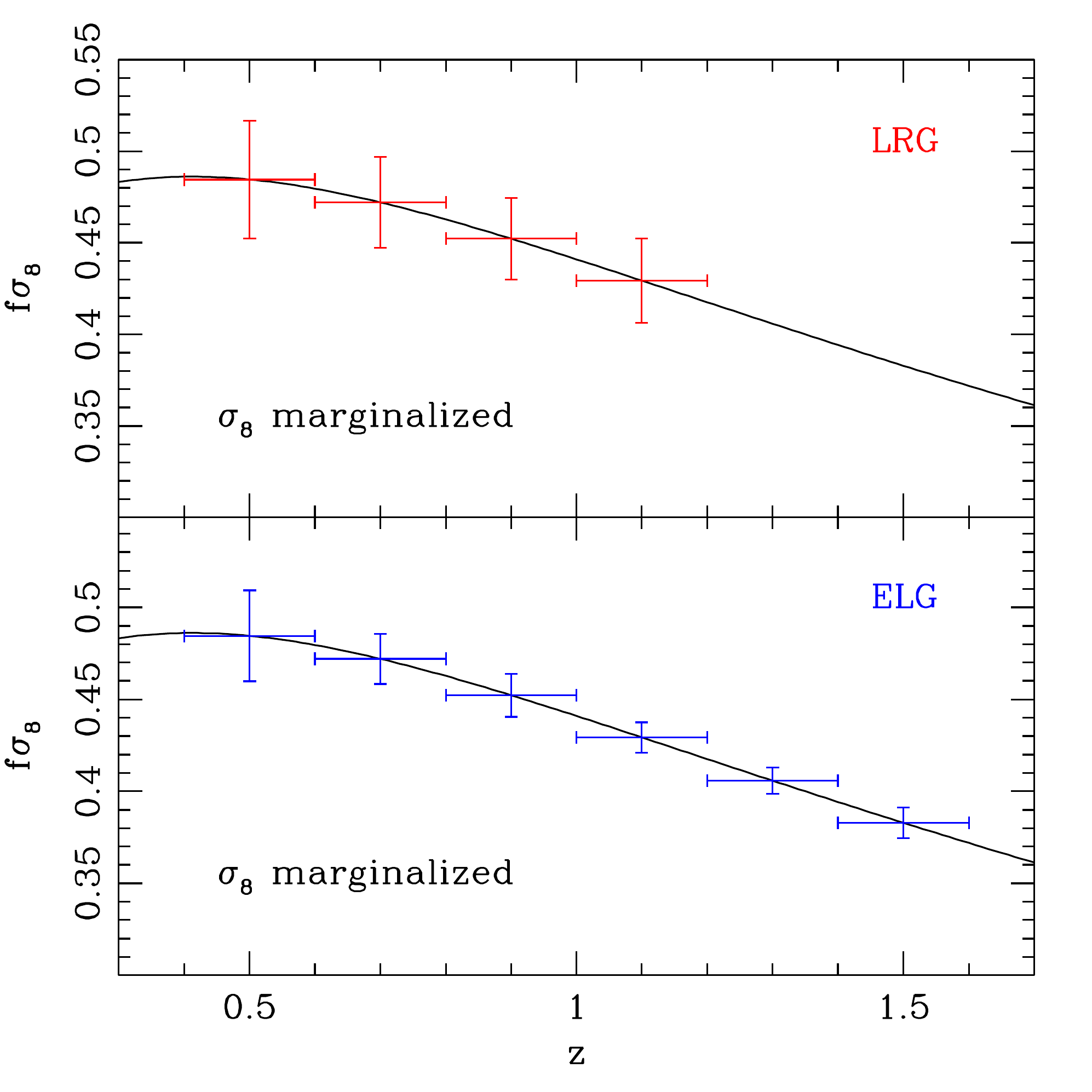}}\hfill
\resizebox{3.in}{!}{\includegraphics{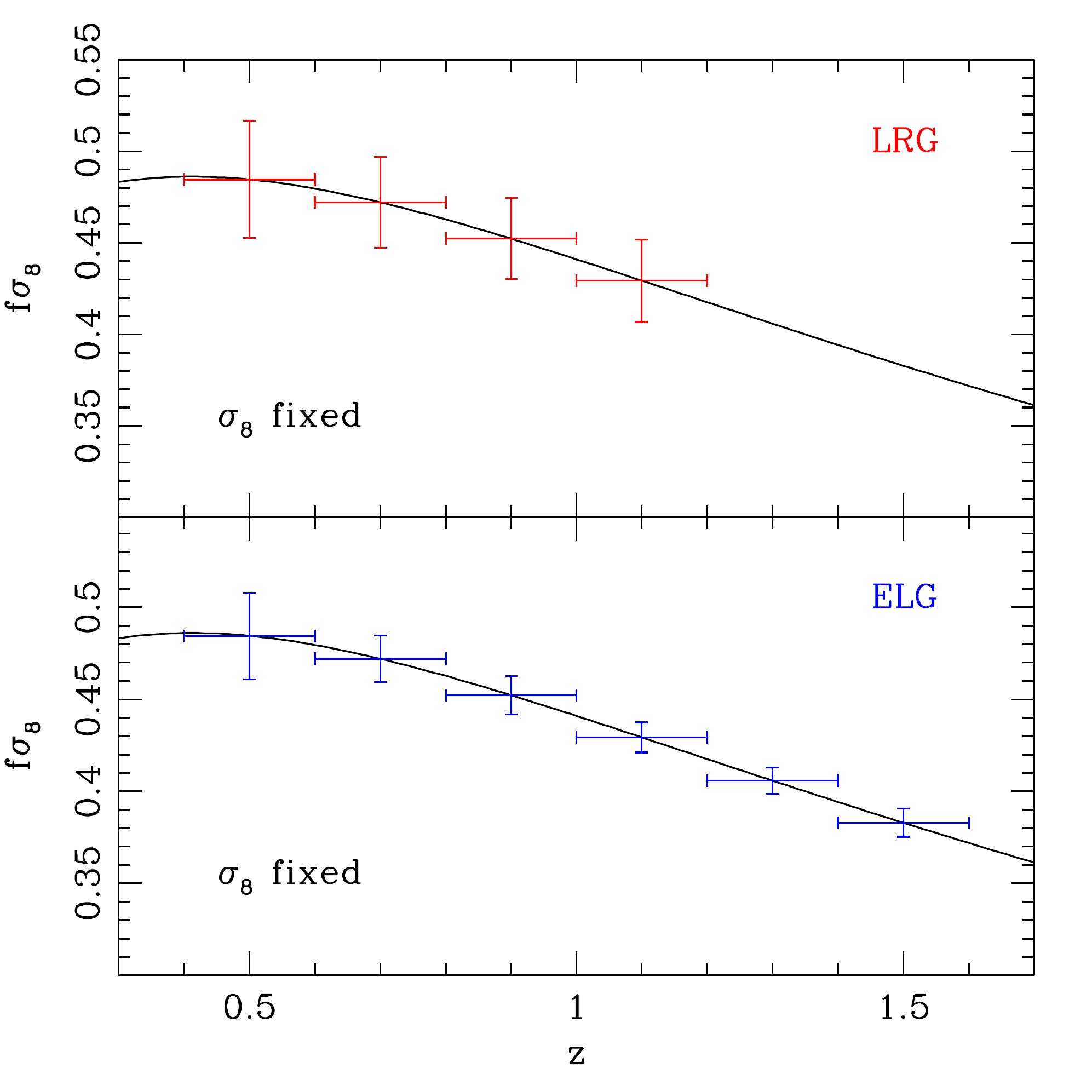}} 
\vspace*{0.5cm}
\end{center}
\caption{{\it Left}: the expected errors on $f\sigma_8$ for DESI, with $\sigma_8$ marginalized over. The result derived from the Fisher matrix, depicted as the error bar around the fiducial value, are plotted against redshift.  The upper and lower panels respectively show the estimated constraints on $f\sigma_8$ derived from the LRG and ELG type galaxies. {\it Right}: Same as in the left panel, but the parameter $\sigma_8$ is fixed, and is not treated as a free parameter. 
\label{fig:fs8}}
\end{figure*}

Incorporating the bias prescription given above into Eq.~(\ref{eq:Pkred_final}), our hybrid RSD model in Sec.~\ref{subsec:hybrid_DM_RSD} and Appendix \ref{appendix:hybrid} enables us to predict the redshift-space galaxy/halo power spectrum beyond the linear regime. To see the behavior of the model beyond the linear bias prescription, left panel of Fig.~\ref{fig:dPdx} shows the logarithmic response of the real-space power spectrum to the quantities $b_1$ and $G_\delta$, depicted as respectively blue and black lines. In linear theory, there is no difference between these responses, which exactly give $2$ (dashed and dotted).   On the other hand, the hybrid model involving the galaxy bias expansion exhibits a wiggle feature for the response to $G_\delta$, showing a small but visible deviation from $2$. This is caused by non--linear smoothing around BAO peaks. On the other hand, the response to $b_1$ still lies at $2$ even in the hybrid model, indicating that the dependence of $G_\delta$ or $\sigma_8$ and $b_1$ becomes distinguishable beyond the linear regime. The selected fiducial values of galaxy biases are presented in Table~\ref{tab:DESI}. The fiducial values for coherent bias $b_1$ are taken from DESI prediction~\cite{DESI16I}, and the redshift dependence of $b_2$ biases are fitting results from halo catalogue.


To feel some flavors on how the degeneracy is broken, left panel of Fig.~\ref{fig:dPdx} presents the response of the redshift-space power spectrum to the variation of $G_\delta$ and $G_\Theta$. The quantity $G_\delta$ is the linear growth factor normalized at primordial epoch as $G_\delta(z_i)(1+z_i)=1$, and is related to $\sigma_8(z)$ through $\sigma_8(z)=[G_\delta(z)/G_\delta(0)]\,\sigma_8(0)$. In left panel of Fig.~\ref{fig:dPdx}, the response of the redshift-space galaxy/halo power spectrum to the variation of $G_\delta$ is shown for both linear theory prediction and hybrid model, depicted as black dotted and solid curves respectively. The function $G_\Theta$ is related to the linear growth rate $f$ through $G_\Theta=f\,G_\delta$. On the other hand, right panel of Fig.~\ref{fig:dPdx} plots the response of the 2D redshift-space power spectrum to the variations of $G_\delta$ and $G_\Theta$ at top and bottom panels respectively, specifically fixing the directional cosine $\mu$ to $0.0$, $0.3$, $0.6$ and $0.9$. The distinct behavior of the power spectrum is observed by varying $\mu$, and the results of hybrid model start to deviate from linear theory predictions, indicating that the parameter degeneracy between $G_\delta$ and $G_\Theta$ is broken, and so is the case for $\sigma_8$ and $f\sigma_8$.

\section{Forecast constraints on $\sigma_8$ and $f\sigma_8$}
\label{sec:Fisher_forecast}

\begin{table}[b]
\begin{ruledtabular}
\begin{tabular}{c|c|c|c|c|c|c|c}
  $z$ & $n_{g}^{\rm LRG}$ & $n_{g}^{\rm ELG}$& $V$ & $b_1^{\rm LRG}$  & $b_1^{\rm ELG}$& $b_2^{\rm LRG}$  & $b_2^{\rm ELG}$\\
  \hline
  0.4--0.6 & $4.9\times 10^{-4}$ & $1.6\times 10^{-4}$ & 3.5 & 2.22 & 1.10 & 0.10 & 0.10\\
  0.6--0.8 & $9.9\times 10^{-4}$ & $1.4\times 10^{-3}$ & 5.4 & 2.45 & 1.21 & 0.67 & 0.67\\
  0.8--1.0 & $3.9\times 10^{-4}$ & $1.7\times 10^{-3}$ & 7.0 & 2.69 & 1.33 & 1.40 & 1.40\\
  1.0--1.2 & $2.4\times 10^{-5}$ & $9.8\times 10^{-4}$ & 8.4 & 2.94 & 1.45 & 2.48 & 2.48\\
  1.2--1.4 & --- & $5.8\times 10^{-4}$ & 9.4 &  --- & 1.57 & --- & 3.91\\
  1.4--1.6 & --- & $2.3\times 10^{-4}$ & 10. &  --- & 1.70 & --- & 5.68\\
\end{tabular}
  \caption{\label{tab:DESI}
The expected number density of LRG and ELG galaxies $n_g^{\rm LRG}[h^3{\rm Mpc}^{-3}]$ and $n_g^{\rm ELG}[h^3{\rm Mpc}^{-3}]$, and survey volume $V[h^{-3}\,{\rm Gpc}^3]$ at each redshift bin used in the Fisher matrix analysis. These specific values are taken from those assumed 
in the DESI experiment \cite{DESI16I}. The fiducial values of bias for LRG and ELG are presented as well, calculated by the formulas $b_{1}^{\rm LRG}(z)D(z) = 1.7$ and $b_{1}^{\rm ELG}(z)D(z) = 0.84$ \cite{DESI16I}, with $D$ being the linear growth factor normalized to unity at present time, i.e., $D(z)=G_\delta(z)/G_\delta(0)$. Concerned with $b_2^{\rm LRG}$ and $b_2^{\rm ELG}$, we exploits the halo fit results. To compute these parameters, Planck $\Lambda$CDM model \cite{PLANK2015} is adopted as the fiducial cosmology. 
\smallskip}
\end{ruledtabular}
\end{table}

\begin{figure*}
\begin{center}
\resizebox{3.in}{!}{\includegraphics{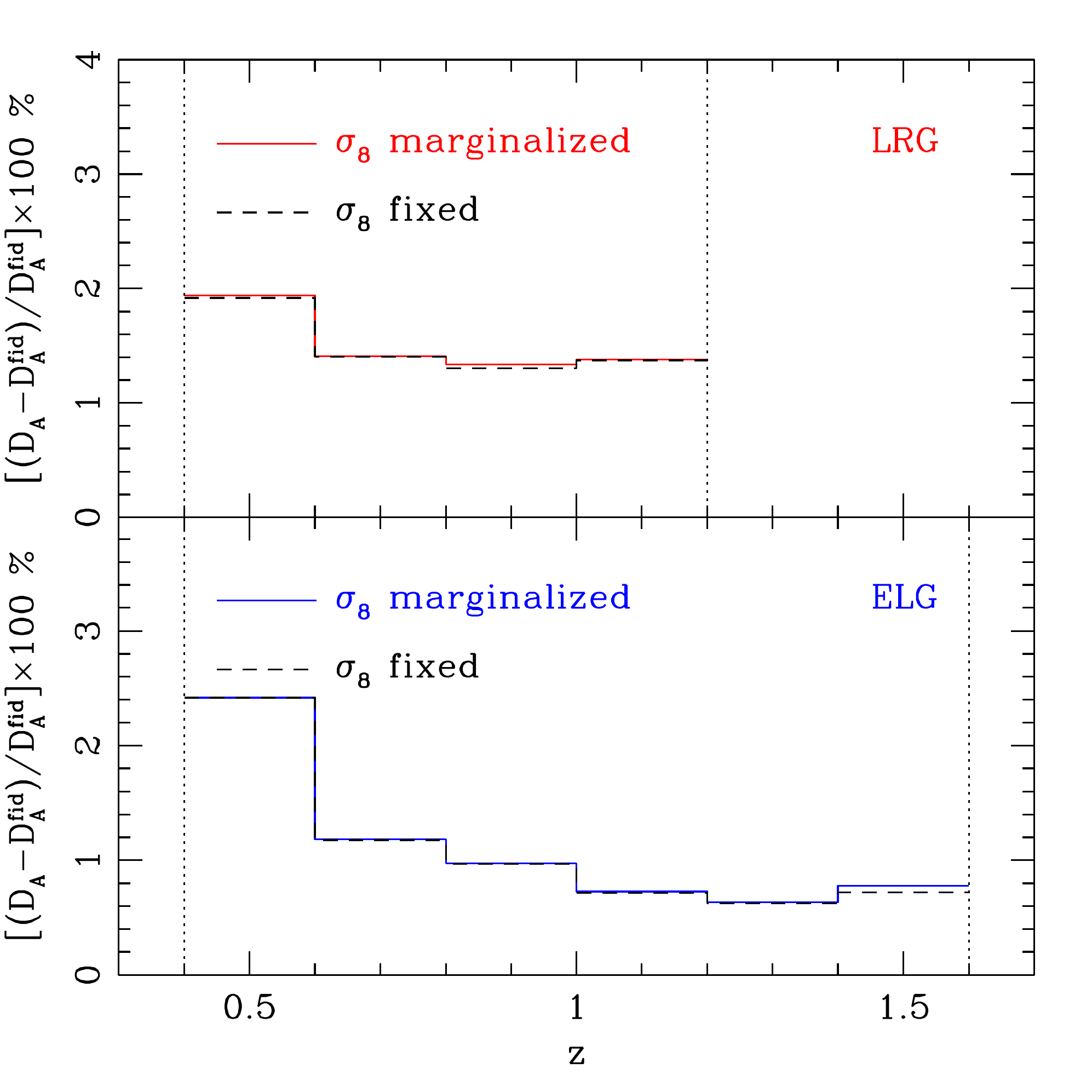}}\hfill
\resizebox{3.in}{!}{\includegraphics{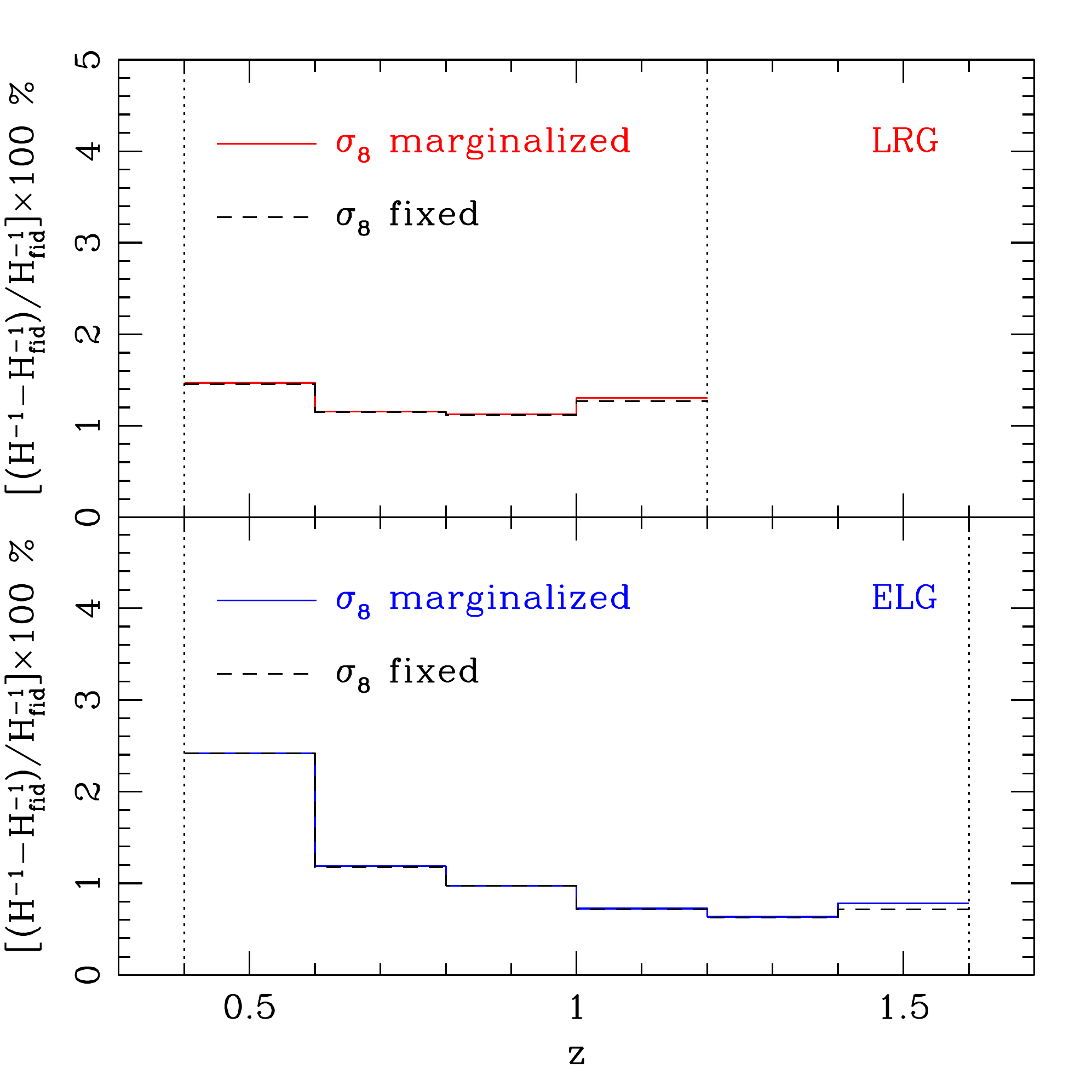}} 
\vspace*{0.5cm}
\end{center}
\caption{{\it Left}: The fractional errors on $D_A$ from DESI LRG (upper) and ELG (lower) samples. The results in units of percentage are plotted as function of redshift. 
 Solid and dashed curves represent the cases with $\sigma_8$ marginalized over and fixed, respectively. {\it Right}: Same as in left panel, but the fractional errors on $H^{-1}$ is shown. }
\label{fig:codis}
\end{figure*}

In this section, based on the model described in Sec.~\ref{sec:hybrid_RSD_model}  as a theoretical template, we demonstrate explicitly how well the degeneracy of the parameter $f\sigma_8$ can be broken, adopting specifically the Dark Energy Spectroscopy Instrument (DESI) \cite{DESI16I}, which is a representative galaxy redshift survey of the so-called stage-IV class~\cite{DETF:2006}, dedicated for a precision measurement of BAO and RSD utilizing the emission-line galaxy (ELG) and luminous red galaxy (LRG) samples (see Table \ref{tab:DESI} for parameter specification, which is discussed below).

\subsection{Fisher matrix formalism}
\label{subsec:Fisher}

To estimate quantitatively the size of the expected errors, we use the Fisher matrix formalism. Regarding the model in Sec.~\ref{sec:hybrid_RSD_model} as an observed power spectrum, the Fisher matrix is evaluated with 
\ba\label{eq:fisher_diag}
F_{\alpha\beta}= \sum_{k,\mu}
\frac{\partial P^{\rm obs}(k,\mu)}{\partial x_{\alpha}}C_{PP}^{-1}\frac{\partial P^{\rm obs}(k,\mu)}{\partial x_{\beta}} 
\ea
where the derivative of the power spectrum is taken with respect to the parameter to estimate, $x_\alpha$. The observed power spectrum $P^{\rm obs}$ is related to $P^{\rm (S)}$ given at Sec.~\ref{sec:hybrid_RSD_model} through the Alcock-Paczynski effect (Ref.~\cite{APtest}, see also Sec.~\ref{sec:intro}):   
\begin{align}
\label{eq:pk_obs_with_AP}
&P^{\rm obs}(k,\mu)=\Bigl(\frac{H}{H_{\rm fid}}\Bigr)
\Bigl(\frac{D_A}{D_{A,{\rm fid}}} \Bigr)^{-2}\,P^{\rm(S)}(q,\nu) \,;\,\,
\\
&\quad
q= \Bigl\{\Bigl(\frac{D_A}{D_{A,{\rm fid}}}\Bigr)^2 (1- \mu^2) + 
\Bigl(\frac{H}{H_{\rm fid}}\Bigr)^{-2} \mu^2\Bigr\}^{1/2}\,k,
\nonumber
\\
&\quad
\nu = \frac{k}{q}\,\Bigl(\frac{H}{H_{\rm fid}}\Bigr)^{-1} \mu. 
\nonumber
\end{align}
In Eq.~(\ref{eq:fisher_diag}), the quantity $C_{PP}$ describes the error covariance of the measured power spectra, whose dominant contributions are the shot noise arising from the discreteness of galaxy distribution, and the cosmic variance due to the limited number of Fourier modes for a finite-volume survey. While the non-Gaussian contribution, leading to the non-zero off-diagonal components, is known to play an important role beyond the linear regime, we will work with the linear Gaussian covariance given below,
\ba
\label{eq:Gaussian_cov}
C_{\rm PP}=\frac{1}{N_P}\left[P^{\rm(S)}(k,\mu) + \frac{1}{n_g^X}\right]^2\,,
\ea
where the quantity $n_g^X$ denotes the number density of galaxies for a specific galaxy type $X$,  i.e., $X=$LRG or ELG for DESI, whose values are summarized in Table \ref{tab:DESI}. The quantity $N_P$ represents the number of available Fourier modes, which is estimated to be
\ba
N_P=\frac{V_{\rm survey}}{2(2\pi)^2} k^2\Delta k\Delta\mu\,
\ea
with $\Delta k$ and $\Delta \mu$ being the bin width of the power spectrum data given in the $(k,\mu)$ plane, with $\Delta k=0.01 h{\rm Mpc^{-1}}$ and $\Delta \mu=0.1$. Here, the quantity $V_{\rm survey}$ is the survey volume, whose values is also listed in Table \ref{tab:DESI}. 

Adapting the Gaussian covariance at Eq.~(\ref{eq:Gaussian_cov}), the forecast results of our Fisher matrix analysis would be optimistic. Nevertheless, at the scales where the perturbative corrections of the next-to-next-to-leading order is still sub-dominant, the impact of non-Gaussian covariance would be small. The forecast results presented below can thus give an important guideline toward a more quantitative analysis. Indeed, to compute the Fisher matrix at Eq.~(\ref{eq:fisher_diag}) below, we will conservatively fix the maximum wavenumber to $k_{\rm max}=0.16\,h$\,Mpc$^{-1}$, at which the non-linearity is still mild at the range of redshifts, $0.4\leq z\leq 1.6$.

To evaluate Eq.~(\ref{eq:fisher_diag}), we will use the power spectrum of $P^{\rm(S)}(k,\mu)$ in Eq.~\ref{eq:Pkred_final}. As the free parameters to estimate, we consider, at each redshift slice, $x_\alpha=\{G_\delta, \,G_\Theta, \, D_A, \, H,\, b_1,b_2,\,\sigma_z\}$, where the former two are simply proportional to $\sigma_8(z)$ and $f\,\sigma_8(z)$. The parameter $\sigma_z$ quantifies the non-perturbative suppression of the power spectrum by the so-called Fingers-of-God damping that appears in the function $D_{\rm FoG}$ at Eq.~(\ref{eq:Pkred_final}). We shall below adopt the Gaussian form of the damping function: 
\bea
D_{\rm FoG}(x)=\exp(-x^2).
\label{eq:D_FoG}
\eea
where the fiducial $\sigma_p=3.3h{\rm Mpc^{-1}}$ computed using linear theory only. Then, the derivative of the power spectrum $P^{\rm obs}$, given at Eq.~(\ref{eq:pk_obs_with_AP}),  is taken with respect to the parameters, $x_\alpha$. Here, the higher-order bias parameters, $b_{s2}$ and $b_{3\rm{nl}}$, are fixed, adopting the following relations: 
\bea
\label{eq:b_s2b_3nl}
b_{s2}=-\frac{4}{7}(b_1-1) \,,
&\quad&
b_{3\rm{nl}}=\frac{32}{315}(b_1-1)\,,\no
\eea
which are derived assuming the Lagrangian local bias, and are shown to be a good approximation for halos and galaxies residing at the halo center (e.g., \cite{Baldauf2012,Chan2012,Saito2014}). The fiducial values of the linear bias parameters for ELG and LRG are presented in Table \ref{tab:DESI}.

\begin{figure}
\begin{center}
\resizebox{3.in}{!}{\includegraphics{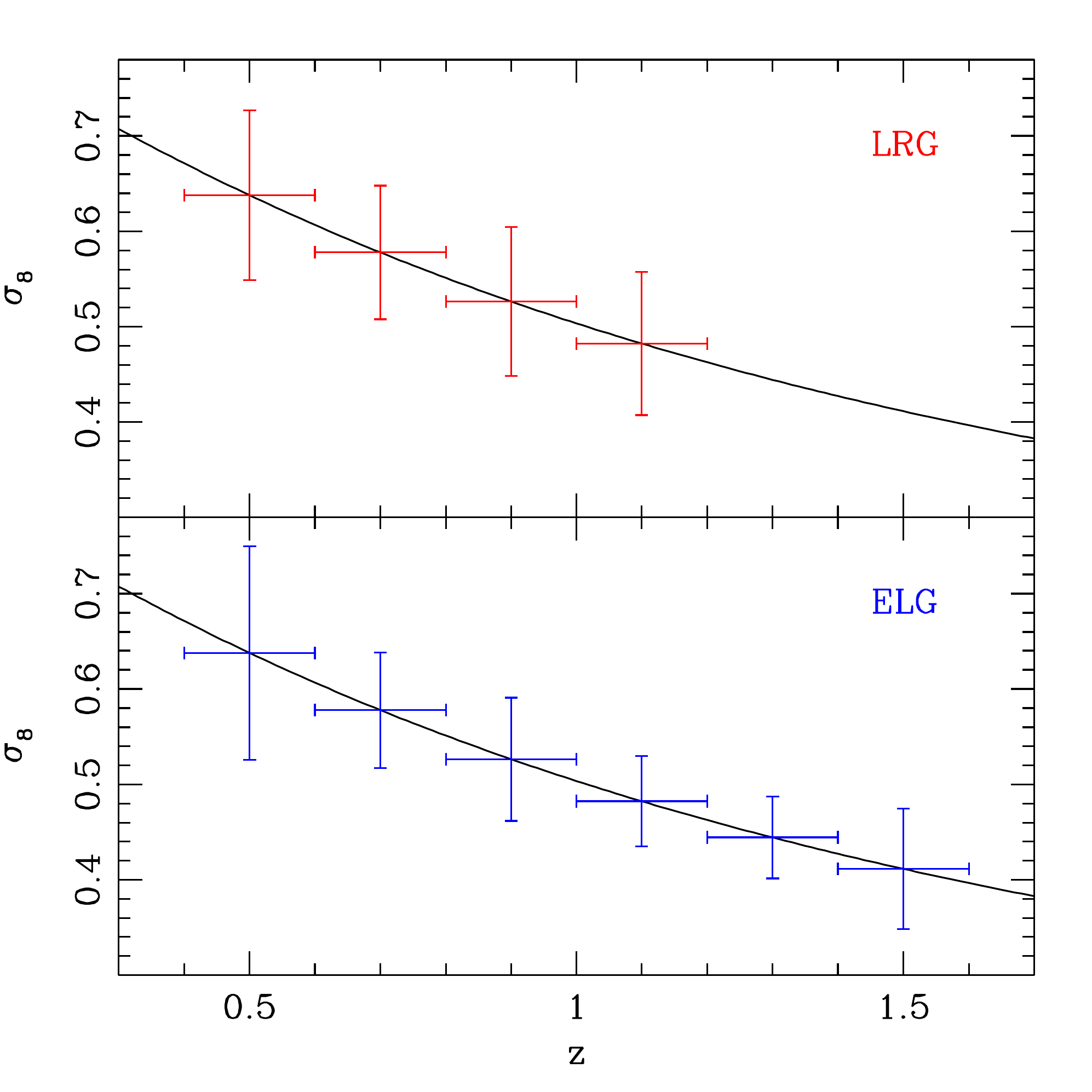}}
\vspace{0.5cm}
\end{center}
\caption{The expected error on $\sigma_8$ for the DESI LRG (upper) and ELG (lower) type samples. }
\label{fig:s8}
\end{figure}

\begin{figure*}
\begin{center}
\resizebox{3.3in}{!}{\includegraphics{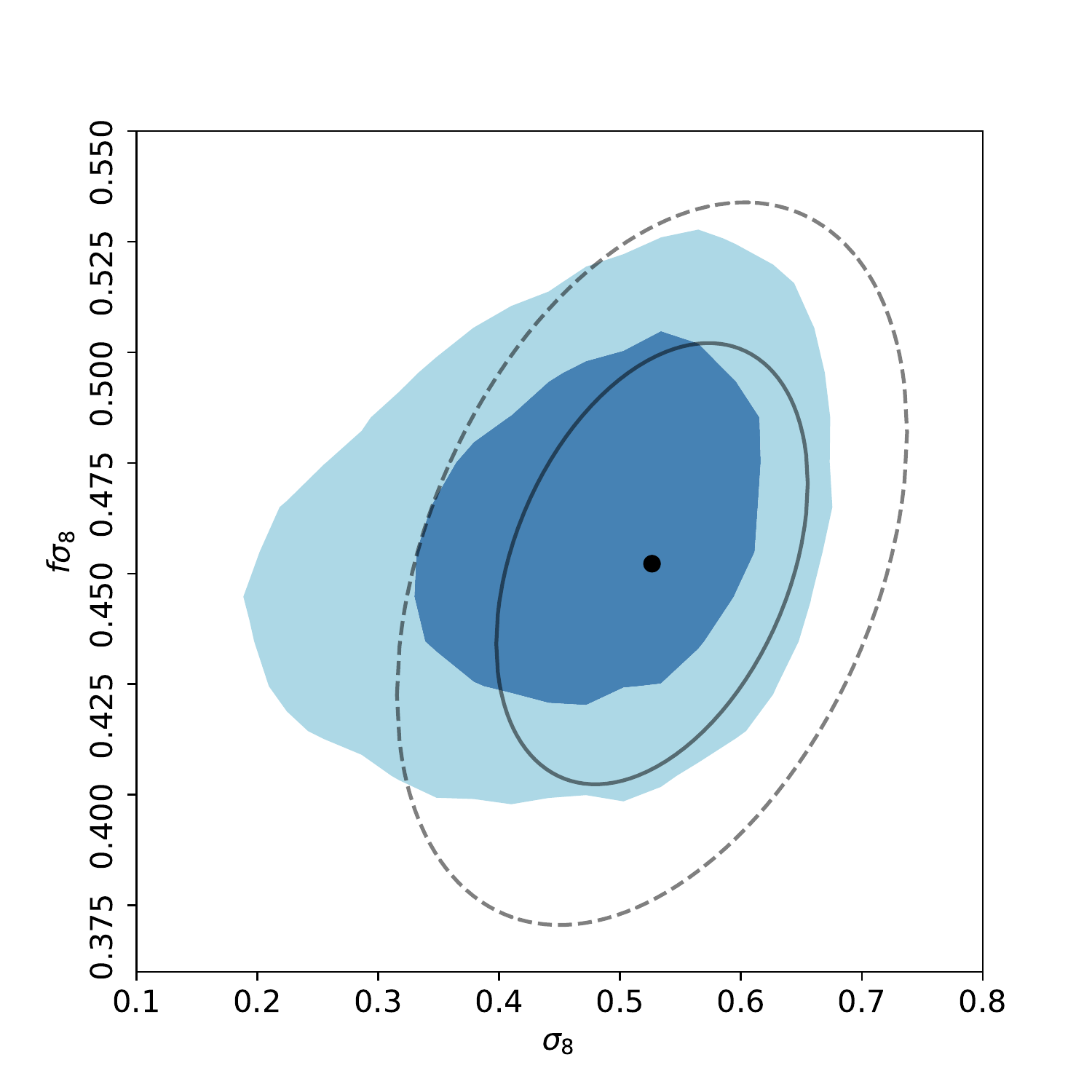}} 
\resizebox{3.3in}{!}{\includegraphics{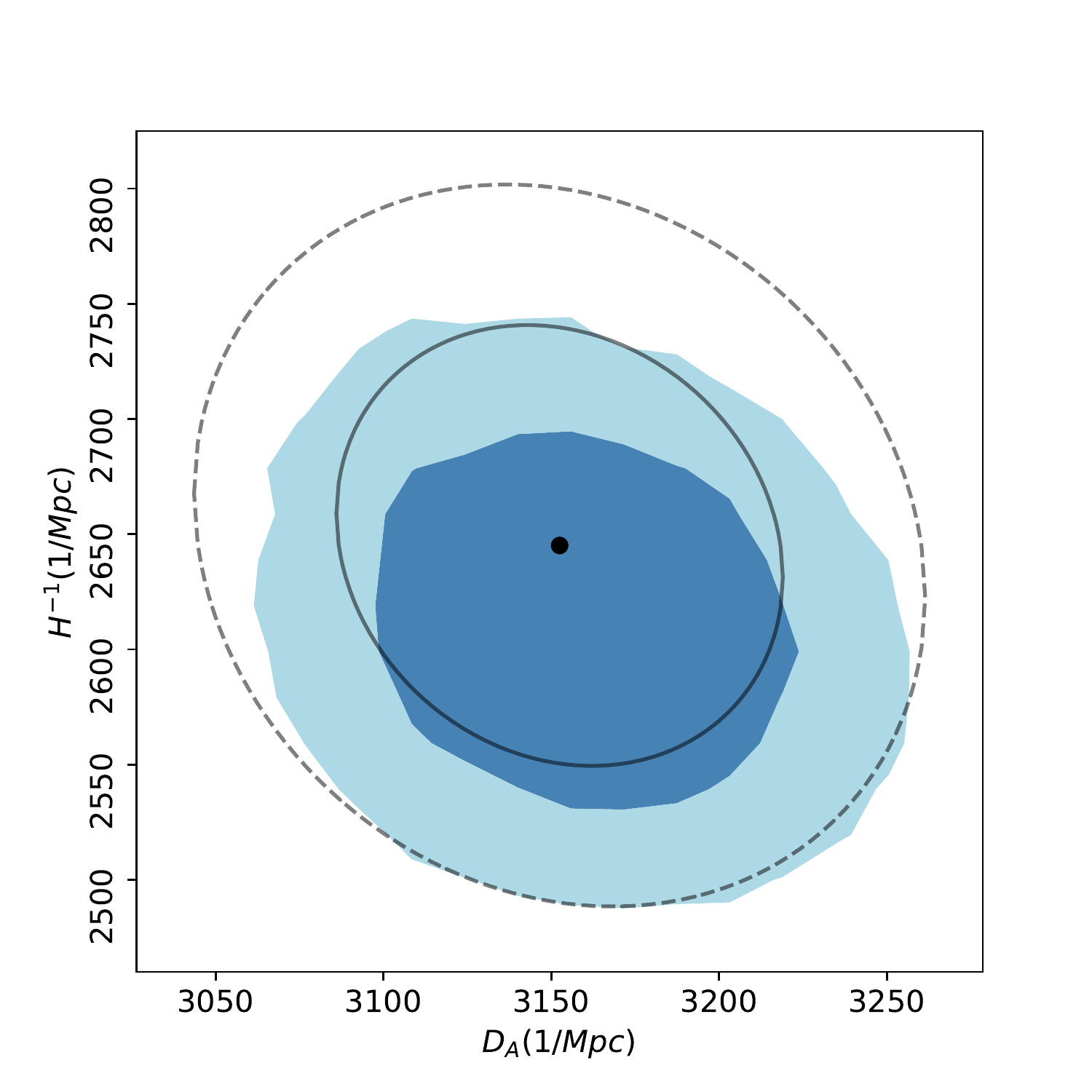}}
\vspace*{0.5cm}
\end{center}
\caption{{\it Left:} The measured and estimated cosmological constraints on $\sigma_8$ and $f\sigma_8$ at 68$\%$ and 95 $\%$ confidence levels are presented with the blue filled and the black unfilled contours respectively. The true values are presented with a black dot in the middle of contours. Those constraints are computed after fully marginalizing other parameters such as ($D_A,H^{-1},\sigma_p, b1, b2$). {\it Right:} The same presentation as for the left panel, but for $D_A$ and $H^{-1}$. Those constraints are computed after fully marginalizing other parameters such as ($\sigma_8,f\sigma_8,\sigma_p, b_1, b_2$).}
\label{fig:MJ}
\end{figure*}

\subsection{Results}
\label{subsec:results}

With the setup described in Sec. ~\ref{subsec:Fisher}, the one-dimensional marginalized errors on the parameters $f\sigma_8$ and $\sigma_8$, as well as the geometric distances, are obtained from the inverse of Fisher matrix, $\Delta x_\alpha=\sqrt{F^{-1}_{\alpha\alpha}}$.

Let us first see how the expected error on the parameter $f\sigma_8$ is altered when the parameter $\sigma_8$ or $G_\delta$ is taken to be free, and is marginalized. Fig.~\ref{fig:fs8} plots the results with $\sigma_8$ marginalized (left) and fixed (right), the latter of which is obtained from the inverse of the sub-matrix $F_{\alpha\beta}$ subtracting the component of $G_\delta$. In each panel, the one-dimensional error on $f\sigma_8$ is shown at each redshift slice for the LRG (top) and ELG (bottom) samples. Remarkably, we found that the size of the error on $f\sigma_8$ almost remains unchanged irrespective of the treatment of the parameter $\sigma_8$, leading to the constraint at the level of a few percent in both ELG and LRG samples, which is expected from the stage-IV class surveys.

To elucidate further the impact of marginalizing the $\sigma_8$ parameter, we also plot in Fig.~\ref{fig:codis} the constraints on the geometric distances. Here, the fractional errors on the angular diameter distances and Hubble parameters , $(D_A-D_{A,{\rm fid}})/D_{A,{\rm fid}}$ and $(H-H_{\rm fid})/H_{\rm fid}$, are respectively shown in left and right panels as function of the redshift. In each panel, the results with $\sigma_8$ marginalized and fixed are depicted as solid and dashed lines. Again, no notable difference is found between the two cases, and even the results marginalizing $\sigma_8$ reach at 1 \% precision for the constraints on both $D_A$ and $H$. In particular, with the ELG samples, the constraint is improved as increasing the redshift out to $z\sim1.5$, with the statistical error down to a sub-percent level.

In Fig.~\ref{fig:s8}, we plot the statistical errors on $\sigma_8$ that is separately determined as a free parameter together with $f\,\sigma_8$. As opposed to the tight constraint on $f\sigma_8$, the $\sigma_8$ appears poorly constrained. Considering the fact that the nonlinear corrections in the power spectrum template that can break the parameter degeneracy are small, this result is reasonable. Rather, accessing the scales to the weakly nonlinear regime, the number of available Fourier modes gets increasing, and the constraints on $f\sigma_8$, as well as geometric distances are improved, as we have seen in Figs.~\ref{fig:fs8} and \ref{fig:codis}. Nevertheless, using the ELG, the expected error on $\sigma_8$ becomes also improved with redshifts, achieving $10$\% precision at $z=1.5$. Since constraining the growth history of structure at higher redshifts particularly helps testing and constraining the gravity as well as the cosmic acceleration, the simultaneous determination of $f\sigma_8$ and $\sigma_8$ is beneficial. Combining the power spectrum with bispectrum, the constraining power will be further improved.

\section{Breaking the $f\,\sigma_8$ degeneracy in $N$-body simulations}
\label{sec:test_MCMC}

Given the survey setup and the theoretical template of power spectrum,  the Fisher matrix analysis in previous section predicts the statistical errors on each free parameter and their parameter degeneracy, but cannot tell how the best-fit  parameters are accurately determined. In this section, to check the validity of the forecast results in Sec.~\ref{sec:Fisher_forecast} as well as to test the hybrid RSD model of power spectrum in Sec.~\ref{sec:hybrid_RSD_model}, we here examine the parameter estimation analysis, based on the Markov chain Monte Carlo (MCMC) technique.

For this purpose, the cosmological $N$-body simulation is carried out by the publicly available code, GADGET-2 \cite{Springel01}, and we ran in total $100$ simulations, with $1024^3$ particles in comoving periodic cubes of the side length $1,890\,h^{-1}$\,Mpc. Using the output results at $z=0.9$, the halo catalog was created using the halo finder, ROCKSTAR~\cite{Rockstar}, with a halo mass range of $10^{13}\,h^{-1}\,M_\odot\,<\,M_h\,<\,10^{13.5}\,h^{-1}\,M_\odot$, and the halo number density $n_h =  1.9\times 10^{-4}\,h^3{\rm Mpc}^{-3}$. The resultant volume of the halo catalog is $6.8\,h^{-3}\,{\rm Gpc}^3$, roughly corresponding to the survey volume of DESI at the redshift slice of $z=0.8--1.0$. The number density of our halo sample is smaller than those expected from DESI LRG and ELG samples (see Table \ref{tab:DESI}), but this does not affect the estimation of statistical errors, as we consider the scales where the shot noise is subdominant. 

The created halo catalog at $z=0.9$ is then used to measure the power spectrum in redshift space, which is compared with the hybrid RSD model to estimate the parameters $(G_\delta,\,G_\Theta,\,D_A,\,H,\,b_1,\, b_2,\,\sigma_z)$, similarly to what has been done in Sec.~\ref{sec:Fisher_forecast}, assuming the Lagrangian local bias. Adopting the same maximum wavenumber as used in Sec.~\ref{sec:Fisher_forecast}, i.e., $k_{\rm max}=0.16\,h$\,Mpc$^{-1}$, the MCMC results marginalized over other parameters are shown in Fig.~\ref{fig:MJ}, where we plot the two-dimensional error contours on $(\sigma_8,\,f\sigma_8)$ (left) and  $(D_A,\,H^{-1})$ (right). The blue shaded regions represent the $68$\% (dark) and $95$\% (light) credible regions obtained from the MCMC analysis, which exhibit roughly the  

In Fig.~\ref{fig:MJ}, we also plot the forecast results for the Fisher matrix analysis. The solid and dashed contours centered at the fiducial parameters, indicated by the black filled circles, are respectively the the $68$\% and $95$\% credible regions for the forecast errors. The MCMC results reproduce well the forecast results of the Fisher matrix analysis, and the $68$\% credible regions consistently include the fiducial parameters. Thus, with the hybrid RSD model presented in Sec.~\ref{sec:hybrid_RSD_model}, unbiased parameter estimation is shown to be possible, with the degeneracy between the parameters $f$ and $\sigma_8$ broken. Although the statistical error of $\sigma_8$ is large, this is the first demonstration that the simultaneous determination of $f$ and $\sigma_8$ is possible only with the power spectrum at weakly nonlinear scales. 



\section{Conclusion}
\label{sec:conclusion}

Redshift-space distortions (RSD) that appear in the observed galaxy distributions via spectroscopic surveys offer an important clue to test the gravity on cosmological scales. Combining the measurement of baryon acoustic osculations (BAO), RSD can be also used to clarify the nature of cosmic acceleration. Toward an unambiguous estimation of cosmological parameters, a crucial issue is not only to improve the precision of RSD measurement, but also to exploit the method to disentangle the parameter degeneracy inherent in the observables. 

In this paper, on the basis of an accurate template for the galaxy/halo redshift-space power spectrum, we get access to the weakly nonlinear regime, and showed that the parameter degeneracy inherent in the linear-theory power spectrum can be broken. To be precise, in linear theory, the linear growth rate $f$ and the fluctuation amplitude $\sigma_8$ appear in the form of $f\sigma_8$ [Eq.~(\ref{eq:linear_pk})], and this degeneracy cannot be broken unless the galaxy bias parameter is a priori known or is accurately determined. In order to break the degeneracy, one way is to go beyond the linear theory. Here, we use the hybrid RSD model of power spectrum that has been developed in our previous papers. Based on the perturbation theory calculations, the model incorporates the higher-order corrections calibrated with $N$-body simulations into the power spectrum expressions, and the accuracy of predictions is improved. Taking further the galaxy bias into account, the hybrid model enables us to get access to the observed galaxy power spectrum at the weakly nonlinear regime. 

Employing the Fisher matrix analysis, we show explicitly that the degeneracy of the parameter $f\sigma_8$ can be broken, and $\sigma_8$ is separately estimated in the presence of galaxy bias. The statistical errors on $f\sigma_8$ as well as the geometric distances $D_A$ and $H$, determined from the BAO via the Alcock-Paczynski effect, are found to remain unchanged, irrespective of whether we treat $\sigma_8$ or the growth factor $D_\delta$ as a free parameter to marginalize or not. As a result, we have shown that the Dark Energy Survey Instrument, as a representative stage-IV class galaxy survey, can unambiguously determine $\sigma_8$ at the precision of $\sim10$\% at higher redshifts even if we restrict the accessible scales to $k\lesssim0.16\,h$\,Mpc$^{-1}$. Further, performing the Markov chain Monte Carlo analysis, we explicitly demonstrate that with the hybrid RSD model, the parameters $f\sigma_8$ and $\sigma_8$ are simultaneously estimated, and their fiducial values can be properly recovered, with the statistical errors fully consistent with the forecast results of the Fisher matrix analysis.

While the analysis in the present paper gives a first explicit demonstration on how well the parameter degeneracy in the measurement of RSD can be broken, the ability to achieve this heavily relies on the theoretical template of the observed power spectrum. Toward a further improvement of the cosmological constraints in an unbiased way, one needs to develop a model that can get access to smaller scales. Though the present paper considered the perturbation theory based model aided by the $N$-body simulations, a simulation based model such as the so-called emulator would be certainly powerful (e.g. \cite{Kobayashi2020}). With such a model, the accessible range of wavenumber becomes broader, and the simultaneous constraints on $f\sigma_8$ and $\sigma_8$, as well as the geometric distances, will be improved in a greater precision. The discussion along the direction of this should be done in the near future.

\section*{Acknowledgments}

We would like to thank Takahiro Nishimichi for useful discussions and comments. Numerical calculations were performed by using a high performance computing cluster in the Korea Astronomy and Space Science Institute. YZ acknowledges the support from the Guangdong Basic and Applied Basic Research Foundation No.2019A1515111098. AT acknowledges the support from MEXT/JSPS KAKENHI Grant No. JP16H03977, JP17H06359 and JP20H05861. AT was also supported by JST AIP Acceleration Research Grant NO. JP20317829, Japan.

\appendix

\section{Description of hybrid RSD model for matter power spectrum}
\label{appendix:hybrid}

The evolution of the density and peculiar velocity fields is traditionally parameterized by $\sigma_8$ and $f\sigma_8$. Both are related to the growth functions of $G_\delta$ and $G_\Theta$ written as below,
\ba
\sigma_8^2(z)=\frac{G_\delta^2(z)}{2\pi^2}\int W_8^2(k) k^2 P^i_{\delta\delta}(k) dk\,,
\ea
\ba
(f\sigma_8)^2(z)=\frac{G_\Theta^2(z)}{2\pi^2}\int W_8^2(k) k^2 P^i_{\delta\delta}(k) dk\,.
\ea
The window function $W_8$ is given by,
\ba
W_8=\frac{3j_1(kR_8)}{kR_8}\,,
\ea
where $j_1$ is the first-order spherical Bessel function, and $R_8=8 h^{-1}{\rm Mpc}$.
We intend to simultaneously probe $\sigma_8$ and $f\sigma_8$ by exploiting the $G_\delta$ and $G_\Theta$ parameters.

In the hybrid approach of modelling the RSD effect, the theoretical spectra $\bar P^{\rm th}_{XY}(k,z)$ $(X,Y=\delta\,\,\mbox{or}\,\,\Theta)$ are computed by the RegPT treatment~\cite{RegPT}, in which all the statistical quantities including power spectrum are expanded in terms of the multi-point propagators up to the two--loop order as,
\ba
&&\bar P_{XY}(k,z) = \bar \Gamma_X^{(1)}(k,z)\bar \Gamma_Y^{(1)}(k,z)\bar P^{i}(k)
\nn
\\
&&\quad
+2\int\frac{d^3\vec q}{(2\pi)^3}\bar \Gamma_X^{(2)}(\vec q, \vec k-\vec q,z) \bar \Gamma_Y^{(2)}(\vec q, \vec k-\vec q,z)
\nn
\\
&&\quad\quad\quad
\times \bar P^i(q) \bar P^{i}(|\vec k-\vec q|) 
\nn
\\
&&\quad
+6\int\frac{d^3\vec pd^3\vec q}{(2\pi)^6}
\nn
\\
&&\quad\quad\quad
\times \bar \Gamma_X^{(3)}(\vec p,\vec q, \vec k-\vec p-\vec q,z)\bar \Gamma_Y^{(3)}(\vec p,\vec q, \vec k-\vec p-\vec q,z) 
\nn
\\
&&\quad\quad\quad
\times\bar P^i(p) \bar P^i(q) \bar P^i(|\vec k-\vec p-\vec q|).
\label{eq:pk_RegPT}
\ea
Here $\bar P^i$ is the initial power spectrum, and $\Gamma_X^{(n)}$ is the $(n+1)$-point propagator. In RegPT treatment, the propagators are constructed with the standard PT calculations. While the standard PT is usually applied to a limited range of wavenumbers, incorporating the result of a partial resummation in the high-$k$ limit, a regularized prediction of the propagators could be applicable to a larger $k$. 

Let us first see the two-point propagator $\bar\Gamma_X^{(1)}$. The expression relevant at the two-loop order is summarized as
\ba\label{eq:gammax1}
\bar \Gamma_X^{(1)}(k,z) = {\rm exp}\left(-\bar G_\delta^2\,\bar\gamma \right)\sum_n \bar G_X \bar G_\delta^{n-1} \bar{\cal C}^{(1)}_n(\bar \gamma).
\ea
Here,  $\bar\gamma$ is defined by $\bar\gamma=k^2\bar \sigma^2_{\rm d}/2$ with $\bar \sigma_{\rm d}^2$ being the dispersion of displacement field. The $\sigma_{\rm d}$ is computed with the initial power spectrum through~\footnote{Note that choice of the upper bound of the integral has been specified in
\cite{RegPT} in somewhat phenomenological way,  and there might be a possible uncertainty. However, it can be absorbed into $\bar{\cal O}$ in our prescription at least perturbatively.} $\bar \sigma_{\rm d}^2=\int_0^{k/2} (dq/6\pi^2)\bar P^i(q)$. The coefficients $\bar{\cal C}^{(n)}$ in Eq.~(\ref{eq:gammax1}) are expressed in terms of the standard PT results, and including the theoretical uncertainties, they are given by
\ba
\bar{\cal C}^{(1)}_1(\bar \gamma) &=&1, \nn\\
\bar{\cal C}^{(1)}_3(\bar \gamma) &=& \bar \gamma + \bar \Gamma_{X,{\rm 1-loop}}^{(1)}(k),\nn\\
\bar{\cal C}^{(1)}_5(\bar \gamma) &=& \bar \gamma^2/2 + \bar \gamma\bar \Gamma_{X,{\rm 1-loop}}^{(1)}(k) + \bar \Gamma_{X,{\rm 2-loop}}^{(1)}(k)+\bar{\cal O}^{(1)}_{X,5},\nn\\
\bar{\cal C}^{(1)}_n(\bar \gamma) &=& \bar{\cal O}^{(1)}_{X,n}\,,
\ea
and $\bar {\cal C}^{(1)}_n=0$ for even numbers of $n$. The $\bar G_X$ denotes the density ($X=\delta$) and velocity ($X=\Theta$) growth functions for the fiducial cosmology at the redshift $z$. These growth functions are the key quantities to be estimated from observations
, and are related to the linear growth factor $D_+$ and linear growth rate $f$ 
through $G_\delta=D_+$ and $G_\theta=f\,D_+$. The function $\Gamma_{X,{\rm n-loop}}^{(p)}$ represents the standard PT $(p+1)$-point propagator at the $n$-loop order, whose explicit expression is given in
~\cite{RegPT,Bernardeau14}. The quantity $\bar{\cal O}^{(1)}_{X,n}$ characterizes the uncertainties or systematics in PT, which will be later calibrated with $N$-body simulations. We assume that the uncertainties arise not only from the higher-order (three-loop) but also from the two-loop order, partly due to the UV sensitive behavior of the single-stream PT calculation.

Similarly, the expression of the three-point propagator, $\bar \Gamma_X^{(2)}(k,z)$ is given by
\ba
\bar \Gamma_X^{(2)}(k,z) &=& {\rm exp}\left(-\bar G_\delta^2\,\bar\gamma \right)\sum_n \bar G_X\,\bar G_\delta^{n-1} \bar{\cal C}^{(2)}_n\,\,;\nn \\
\bar{\cal C}^{(2)}_2(\bar \gamma) &=& \bar F_X^{(2)}(\vec q, \vec k-\vec q),\nn\\
\bar{\cal C}^{(2)}_4(\bar \gamma) &=& \frac{\bar\gamma}{2} \bar F_X^{(2)}(\vec q, \vec k-\vec q)+\bar\Gamma_{X,{\rm 1-loop}}^{(2)}(\vec q, \vec k-\vec q)+\bar{\cal O}^{(2)}_{X,4}\,, \nn\\ 
\bar{\cal C}^{(2)}_n(\bar \gamma) &=& \bar{\cal O}^{(2)}_{X,n}\,,
\ea
and $\bar{\cal C}^{(2)}_n=0$ for odd number of $n$. Also, the expression of the four-point propagator, $\bar \Gamma_X^{(3)}(k,z)$, relevant at the two-loop order, is
\ba
\bar \Gamma_X^{(3)}(k,z) &=& {\rm exp}\left(-\bar G_\delta^2\,\gamma \right)\sum_n \bar G_X\,\bar G_\delta^{n-1} \bar{\cal C}^{(3)}_n\,\,; \\
\bar{\cal C}^{(3)}_3(\bar \gamma)&=&\bar F_X^{(3)}(\vec p,\vec q, \vec k-\vec p-\vec q) + \bar{\cal O}^{(3)}_{X,3}, \\
\bar{\cal C}^{(3)}_n(\bar \gamma) &=& \bar{\cal O}^{(3)}_{X,n}.
\ea
Note that $\bar{\cal O}^{(2)}$ and $\bar{\cal O}^{(3)}$ represent the possible uncertainties.

While the prescription given above is supposed to give an accurate theoretical prediction at higher redshifts and larger scales, Ref.~\cite{Song2018} reported that the RegPT prediction of $\bar{P}_{XY}$ at a low redshift (to be precise $z=0.5$) exhibits a small deviation from $N$-body simulations at $k>0.1\hompc$~. Although this might be partly ascribed to the systematics in the $N$-body simulations, a lack of higher-order terms as well as a small systematics in the PT calculations is known to sensitively affect the high-$k$ prediction. Here, we characterize the difference between the measured and predicted power spectra by $\bar P_{XY}^{\rm res}$. We then divide the power spectrum into two pieces:
\ba
\bar P_{XY}(k,z)=\bar P^{\rm th}_{XY}(k,z)+\bar P^{\rm res}_{XY}(k,z),
\label{eq:pk_hybrid_fiducial}
\ea
where $\bar P^{\rm th}_{XY}$ represents the PT prediction with $\bar{\cal O}^{(m)}_{X,n}\rightarrow 0$. Collecting all the uncertainties introduced in the multi-point propagators, the residual power spectrum $\bar P_{XY}^{\rm res}$ is schematically expressed as
\ba
&&\bar P^{\rm res}_{XY}=\bar G_X \bar G_Y \bar G_\delta^4\,\Biggl\{
\left[{\cal O}^{(1)}_{Y,5} + {\rm higher} \right]\bar P^{i}\nn\\
&&+\left[\bar{\cal O}^{(1)}_{X,5} + {\rm higher} \right]\bar P^{i}
+\int \left[\bar{\cal O}^{(2)}_{Y,4}\bar F_Y^{(2)}+ {\rm higher}\right]\bar P^{i}\bar P^{i}\nn\\
&&+\int \left[\bar{\cal O}^{(2)}_{X,4}\bar F_X^{(2)}+ {\rm higher}\right]\bar P^{i}\bar P^{i} \nn\\
&&+\int\int \left[ \bar{\cal O}^{(3)}_{Y,3}\bar F_Y^{(3)}+ {\rm higher}\right]\bar P^{i}\bar P^{i}\bar P^{i} \nn\\
&&+\bar \int\int \left[\bar{\cal O}^{(3)}_{X,3}\bar F_X^{(3)}+ {\rm higher}\right]\bar P^{i}\bar P^{i}\bar P^{i} \Biggr\}.
\label{eq:P_res_fid}
\ea
Here, the uncertainty $\bar{\cal O}^{(m)}_{X,n}$ is assumed to be small, and to be perturbatively treated. The expression implies that apart from a detailed scale-dependent behavior, time dependence is characterized by 
$G_XG_YG_\delta^4$. Thus, once we calibrate the $\bar P_{XY}^{\rm res}$ at a given redshift, we may use it for the prediction at another redshift by simply rescaling the calibrated residuals. Furthermore, for the cosmological models close to the fiducial model, the scale dependence of the higher-order PT corrections is generally insensitive to the cosmology, and we may also apply the calibrated $\bar P_{XY}^{\rm res}$ to other cosmological models.

Next we introduce the way to calculate the dark matter higher order terms. They are incorporated with Eq. (\ref{eq:higherorder_b1_A}-\ref{eq:higherorder_b1_T}) to calculate the halo higher order terms. To begin with, let us consider the $A$ term. From the explicit form, the $A$ term is divided into six pieces. Here, we specifically write down the expressions in fiducial cosmological model: 
\begin{eqnarray}
  \bar A(k,\mu)
  &=& j_1\,\int d^3\bfx \,\,e^{i\bfk\cdot\bfx}\,\,\langle A_1A_2A_3\rangle_c\nonumber\\
  &=& \sum_{n=1}^{6} \bar {\cal A}_n
\label{eq:A_term}
\end{eqnarray}
Note again that the barred quantities indicate those computed/measured in fiducial cosmological model. The explicit form of $\bar {\cal A}_n$ is given below:
\ba
\bar {\cal A}_1&=& 2j_1\,\int d^3\bfx \,\,e^{i\bfk\cdot\bfx}\,\,\langle u_z(\bfr) \delta(\bfr) \delta(\bfr')\rangle_c, 
\label{eq:A_1}\\
\bar {\cal A}_2  &=& j_1\,\int d^3\bfx \,\,e^{i\bfk\cdot\bfx}\,\,\langle u_z(\bfr)\delta(\bfr)\,\nabla_zu_z(\bfr')\rangle_c, 
\label{eq:A_2}\\
\bar {\cal A}_3  &=& j_1\,\int d^3\bfx \,\,e^{i\bfk\cdot\bfx}\,\,\langle u_z(\bfr)\,\nabla_zu_z(\bfr)\delta(\bfr')\rangle_c, 
\label{eq:A_3}\\
\bar {\cal A}_4 &=& 2j_1\,\int d^3\bfx \,\,e^{i\bfk\cdot\bfx}\,\,\langle u_z(\bfr)\,\nabla_zu_z(\bfr)\,\nabla_zu_z(\bfr')\rangle_c,
\label{eq:A_4}\\
\bar {\cal A}_5  &=& j_1\,\int d^3\bfx \,\,e^{i\bfk\cdot\bfx}\,\,\langle -\delta(\bfr)u_z(\bfr')\,\nabla_zu_z(\bfr')\rangle_c, 
\label{eq:A_5}\\
\bar {\cal A}_6  &=& j_1\,\int d^3\bfx \,\,e^{i\bfk\cdot\bfx}\,\,\langle -\,\nabla_zu_z(\bfr)u_z(\bfr')\delta(\bfr')\rangle_c.
\label{eq:A_6}
\ea
These terms are measured from $N$-body simulations according to Ref.~\cite{Zheng16a}. To apply the measured results to the prediction in other cosmological models, we assume the scaling ansatz, as similarly adopted in the prediction of power spectrum $P_{\rm XY}$. That is, assuming that the scale-dependence of each term is insensitive to the cosmology, the prediction of each term is made by simply rescaling the measured results. The proposition made here is that the time-dependence of each term is approximately determined by the leading-order growth factor dependence of $u_z$ and $\delta$. Then, $A$ term in general cosmological model is expressed as
\ba\label{eq:estimatedAn}
&&A(k,\mu)
= \sum_{n=1}^{6} {\cal A}_n \\
&&=\left(G_\delta/\bar G_\delta\right)^2\left(G_\Theta/\bar G_\Theta\right)
\bar{\cal A}_1
+\left(G_\delta/\bar G_\delta\right)\left(G_\Theta/\bar G_\Theta\right)^2
\bar{\cal A}_2\nn\\
&&+\left(G_\delta/\bar G_\delta\right)\left(G_\Theta/\bar G_\Theta\right)^2
\bar{\cal A}_3
+\left(G_\Theta/\bar G_\Theta\right)^3
\bar{\cal A}_4\nn\\
&&+\left(G_\delta/\bar G_\delta\right)\left(G_\Theta/\bar G_\Theta\right)^2
\bar{\cal A}_5
+\left(G_\delta/\bar G_\delta\right)\left(G_\Theta/\bar G_\Theta\right)^2
\bar{\cal A}_6\nn
\ea

We then apply the same strategy to other higher-order corrections, $B$, $F$ and $T$. Dividing these corrections into several pieces, the scaling ansatz leads to the following predictions:
\ba\label{eq:estimatedBn}
&&B(k,\mu) = \sum_{n=1}^{4} {\cal B}_n 
\\
&&= \left(G_\delta/\bar G_\delta\right)^2\left(G_\Theta/\bar G_\Theta\right)^2
\bar{\cal B}_1
+\left(G_\delta/\bar G_\delta\right)\left(G_\Theta/\bar G_\Theta\right)^3
\bar{\cal B}_2
\nn 
\\
&&+\left(G_\delta/\bar G_\delta\right)\left(G_\Theta/\bar G_\Theta\right)^3
\bar{\cal B}_3
+\left(G_\Theta/\bar G_\Theta\right)^4
\bar{\cal B}_4
\nn
\\
&&F(k,\mu) = \sum_{n=1}^{3} {\cal F}_n 
\label{eq:estimatedFn}
\\
&&= \left(G_\delta/\bar G_\delta\right)^2\left(G_\Theta/\bar G_\Theta\right)^2
\bar{\cal F}_1
+\left(G_\delta/\bar G_\delta\right)\left(G_\Theta/\bar G_\Theta\right)^3
\bar{\cal F}_2\nn\\
&&+\left(G_\Theta/\bar G_\Theta\right)^4
\bar{\cal F}_3
\nn
\\
&&T(k,\mu)= \sum_{n=1}^{7} {\cal T}_n 
\label{eq:estimatedTn}
\\
&&= \left(G_\delta/\bar G_\delta\right)^2\left(G_\Theta/\bar G_\Theta\right)^2
\bar{\cal T}_1
+\left(G_\delta/\bar G_\delta\right)\left(G_\Theta/\bar G_\Theta\right)^3
\bar{\cal T}_2\nn\\
&&+\left(G_\delta/\bar G_\delta\right)\left(G_\Theta/\bar G_\Theta\right)^3
\bar{\cal T}_3
+\left(G_\Theta/\bar G_\Theta\right)^4
\bar{\cal T}_4
\nn
\\
&&
+\left(G_\delta/\bar G_\delta\right)^2\left(G_\Theta/\bar G_\Theta\right)^2
\bar{\cal T}_5
+\left(G_\delta/\bar G_\delta\right)\left(G_\Theta/\bar G_\Theta\right)^3
\bar{\cal T}_6
\nn
\\
&&
+\left(G_\Theta/\bar G_\Theta\right)^4
\bar{\cal T}_7 
\nn
\ea
Here, the quantities $\bar{\cal B}_n$, $\bar{\cal F}_n$, and  $\bar{\cal T}_n$ are measured in the fiducial cosmological model. Both $\sigma_8$ and $f\sigma_8$ are estimated from the measured growth functions of $G_\delta$ and $G_\Theta$.


%

\end{document}